\documentstyle[12pt]{article}
\def\ccr{\cr\noalign{\medskip}} %
\def\and{{\qquad\hbox{and}\qquad}}
\def\where{{\qquad\hbox{where}\qquad}}

\def\p{{\partial}}
\def\RR{{\rm R\hspace*{-2ex} I\hspace*{1.2ex}}}
\def\o{{\rm o}}
\def\O{{\rm O}}

\def\parag{\hfil\break} 
\def\kikezd{\parag\underbar}
\def\const{{\rm const}}
\def\smallcirc{{\raise 0.5pt \hbox{$\scriptstyle\circ$}}}
\def\smallover#1/#2{\hbox{$\textstyle{#1\over#2}$}} %
\def\2{{\smallover 1/2}}
\def\T{{\cal T}}
\def\L{{\cal L}}

\let\ssection=\section
\renewcommand{\section}{\setcounter{equation}{0}\ssection}

\title{Field--dependent symmetries of a
non-relativistic fluid model}

\bigskip
\author{
M.~Hassa\"{\i}ne and P.~A.~Horv\'athy
\\
Laboratoire de Math\'ematiques et de Physique Th\'eorique\\
Universit\'e de Tours\\
Parc de Grandmont\\
F-37200 TOURS (France)
}
\date{\today}

\begin{document}

\maketitle

\begin{abstract}
As found by Bordemann and Hoppe
and by Jevicki,
a certain non--relativistic
model of an irrotational and isentropic fluid,
related to membranes and to partons, admits
a Poincar\'e symmetry.
Bazeia and Jackiw
associate this dynamical  symmetry to a novel type of
``field--dependent''  action on space--time.
The  ``Kaluza-Klein type'' framework of Duval et al.
is used to explain the origin
of these symmetries and to derive the associated
conserved quantities. In the non-interacting
case, the symmetry extends to the entire conformal group.
\end{abstract}
\vskip10mm
\centerline{\bf Revised version}
\noindent
Key words~: membrane theory, fluid mechanics,
symmetries and conservation laws.

\vfill\eject
\section{Introduction}

The reduction of membrane theory can lead to a simple model,
describing an isentropic, irrotational fluid \cite{BH}.
A similar system can be obtained, e. g.,
by dimensional reduction of relativistic field theory
 \cite{JEV},
and also in the hydrodynamical formulation
of the (non--linear) Schr\"odinger equation \cite{BJ}.
The model, also used in gas dynamics,
 was further discussed by
Bazeia, Jackiw, and Polychronakos
\cite{BJ}, \cite{JP1}, \cite{BAZ}, \cite{JP2};
note also \cite{NUTKU}.

Let us consider the action
\begin{equation}
{\cal S}=\int\!dxdt\underbrace{\Big[- R\p_{t}\Theta
-\frac{1}{2} R(\partial_{x}\Theta)^2}_{{\cal L}_{0}}
-V( R)\Big],
\label{BJ}
\end{equation}
where $ R(x,t)\geq0$ and $\Theta(x,t)$ are real fields
and $V(R)$ is some potential. (Our Lagrange density
differs from that of Bazeia and Jackiw in \cite{BJ}
in a surface term;
the two expressions are hence equivalent.
Albeit similar results hold in any dimension, we shall
restrict ourselves, for simplicity, to $(1+1)$ dimensionsal
space--time, parametrized by position and time, $x$ and $t$.).

The associated Euler-Lagrange equations read
\begin{equation}
\partial_{t} R+\partial_{x}\big( R\partial_{x}\Theta\big)=0,
\qquad
\partial_t\Theta+\frac{1}{2}(\partial_x\Theta)^2=-\frac{dV}{dR}.
\label{eqmotion}
\end{equation}

In what follows, we shall (except in Section 7), restrict
ourselves
to potentials of the form
$
V=cR^\omega,
$
where $c$ and $\omega$ are real constants.
 In the membrane case the effective potential is in particular
\begin{equation}
V(R)=\frac{c}{R},
\qquad
c=\const.
\label{membpot}
\end{equation}

The Lagrangian (\ref{BJ}) is first-order in the time derivative;
it admits therefore an (extended)
Galilean symmetry, with conserved quantities
\begin{equation}
\begin{array}{cc}
H=\displaystyle\int  dx\underbrace{\left(
\frac{1}{2} R(\partial_{x}\Theta)^2+V(R)\right)
}_{{\cal H}}\qquad\hfill
&\hbox{energy}\\
\cr
P=\displaystyle\int\! dx\underbrace{
 R\,\partial_{x}\Theta}
_{{\cal P}}\hfill
&\hbox{momentum}\\
\cr
B=\displaystyle\int\! dx\,\big(x R-t{\cal P}\big)\hfill
&\hbox{boosts}\\
\cr
N=\displaystyle\int\! dx\, R\qquad\hfill
&\hbox{particle number}\\
\end{array}
\label{Galconst}
\end{equation}

Unexpectedly, the free and the membrane systems
both admit two additional conserved quantities
\cite{BH}, \cite{JEV} \cite{BJ}, namely
\begin{equation}
\begin{array}{cc}
G=\displaystyle\int dx\,
\big(x{\cal H}-\Theta{\cal P}\big)\hfill
&\hbox{``antiboost''}\qquad\hfill
\\
\cr
D=tH-\displaystyle\int dx\, R\Theta\qquad\hfill
&\hbox{time dilatation}\hfill
\hfill
\label{BJconst}
\end{array}
\end{equation}

The generators
(\ref{Galconst})--(\ref{BJconst}) span furthermore
the  $(2+1)$ dimensional Poincar\'e algebra \cite{BH},
 \cite{JEV}, \cite{BJ}.
The arisal of the typically {\it relativistic}
Poincar\'e symmetry for a {\it non-relativistic system}
is quite surprising.
 The mystery is increased by that this
symmetry is {\it not} associated
to any finite--dimensional group action on space-time.
It belongs in fact
 to a new type of ``field-dependent'' non-linear
action on space--time \cite{BJ} which, to our knowledge,
 has never been met before.

Before explaining how these symmetries arise, we point out
that, in the ``free case'' $V=0$,
the entire conformal group $\O(3,2)$ is a symmetry;
it is reduced to the Poincar\'e group for
$V(R)=c/R$, and to the Schr\"odinger group
(the symmetry of the free Schr\"odinger equation \cite{JHN})
 for $V(R)=cR^3$, respectively.

\goodbreak
Where do these symmetries come from~?
We answer this question
by unfolding the system into a higher--dimensional space, obtained
by promoting the ``phase'' $\Theta$  to a ``vertical'' coordinate
(we denote by $s$) on extended space, $M$.
Such a ``Kaluza--Klein--type''  framework for non--relativistic
physics was put forward by Duval et al. \cite{barg}.
In our case, their extended space
$M$ is  $(2+1)$--dimensional Minkowski space,
with $x$ a spacelike and $t$ and $s$ light--cone coordinates.
Then the strange, field--dependent, non--linear action
of Bazeia and Jackiw \cite{BJ}, (Eq. (\ref{BJtransf}) below),
becomes the natural, linear action of the $(2+1)$--dimensional
Poincar\'e group on extended space.

Our starting point is the simple but crucial observation due
to Christian Duval \cite{Duval} which says that, on extended space
$M$, the ``antiboosts'' are
the counterparts of galilean boosts, when galilean time, $t$, and
the ``vertical coordinate'', $s$ are interchanged,
\begin{equation}
t\longleftrightarrow s.
\end{equation}

Many results presented in this paper come by exploiting
this intechange--symmetry.
For example,

\vskip2mm
{$\bullet$} applied to the Galilei group,
the Poincar\'e group is obtained;

{$\bullet$} applied to ``non-relativistic
conformal symmetries'' (Eq. (\ref{nrctransf}) below) yields
relativistic
conformal symmetries,

\vskip2mm\noindent
etc.
It also provides a clue for the non--conventional
implementations on fields.

\goodbreak
The action of the conformal group $\O(3,2)$ and its various
subgroups on $M$ is presented in Section 4. In Section 5
we project the natural, linear action on extended space to
a ``field-dependent action'' on ordinary space. This requires to
generalise as in Eq. (\ref{newequiv}) the
usual equivariance condition (\ref{oldequiv}) of Duval et
al. \cite{barg}.
The authors of Ref. \cite{BJ}
call the Poincar\'e symmetry ``dynamical''
since it is not associated to a natural ``geometric''
action on space--time. Our point is that these symmetries become
``geometric'' on extended space.

In Section 6 we study physics in the extended space
and show how the previous results can be recovered.
Our results show also that the
``membrane potential''(\ref{membpot}) i.e. $V(R)=c/R$ is
the only one which
can accomodate these new type of symmetries.
This is the reason why these strange
symmetries do {\it not} arise for the ordinary Schr\"odinger
equation~:
this latter corresponds in fact to a particular effective
potential, namely to
$
\overline{V}=-\frac{1}{8}
\frac{(\vec{\nabla} R)^2}{R}.
$
Usual equivariance allows us in turn to recover the
well-known Schr\"odinger symmetry.

In Ref. \cite{JP2}, the Poincar\'e symmetry of the fluid system
(\ref{BJ}) is related to that of the Nambu-Goto action of a
membrane
moving in higher dimensional space-time.
Our ``Kaluza--Klein'' framework is an
alternative way of obtaining the same conclusion.

\goodbreak
\section{Symmetries}

We first recall the construction of the conserved quantities.
Let us consider a non-relativistic theory given by the Lagrange
density $\L(\partial_{\alpha}\phi,\phi)$, where $\phi$ denotes
all fields collectively. Then Noether's theorem \cite{JM} says that
if the Lagrange density changes by a surface term
under an infinitesimal transformation
$\phi\to\phi+\delta\phi$,
\begin{equation}
\delta{\cal L}=\p_{\alpha}C^\alpha,
\end{equation}
then
$
J^\alpha=\frac{\delta\L}{\delta(\p_{\alpha}\phi)}
\delta\phi-C^\alpha
$
is a conserved current, $\p_{\alpha}J^\alpha=0$, so that
\begin{equation}
\displaystyle{\int{\!dx\,
{\Big(\frac{\delta\L}{\delta(\p_{t}\phi)}\delta\phi-C^t\Big)}}}
\label{consquant}
\end{equation}
is conserved.
For example, the usual Galilean
 transformations of non--relati\-vis\-tic space-time,
$\left(\begin{array}{c}
x
\\
t
\end{array}\right)
\to
\left(\begin{array}{c}
{x}^\star
\\
{t}^\star
\end{array}\right)
$,
$\left(\begin{array}{c}
R(x,t)
\\
\Theta(x,t)
\end{array}\right)
\to
\left(\begin{array}{c}
{R}^\star(x,t)
\\
{\Theta}^\star(x,t)
\end{array}\right)
$,
where
\begin{equation}
\begin{array}{ccc}
\begin{array}{c}
{x}^\star=x,\hfill
\\
{t}^\star=t+\tau,\hfill
\\
\end{array}\hfill
&
\begin{array}{c}
R^\star(x,t)=R(x,t+\tau),\hfill
\\
\Theta^\star(x,t)=\Theta(x,t+\tau);\hfill\\
\end{array}\hfill
&\hbox{time translation}
\\
\cr
\begin{array}{c}
x^\star=x-\gamma,\hfill
\\
{t}^\star=t,\hfill
\\
\end{array}\hfill
&\begin{array}{c}
R^\star(x,t)= R(x-\gamma,t),\hfill
\\
\Theta^\star(x,t)=\Theta(x-\gamma,t);\hfill\\
\end{array}\hfill
&\hbox{
translation}\\
\cr
\begin{array}{c}
{x}^\star=x+\beta t,
\hfill
\\
{t}^\star=t,\hfill
\\
\\
\end{array}\hfill
&\begin{array}{cc}
R^\star(x,t)=& R(x+\beta t,t),\hfill
\\
\Theta^\star(x,t)=&
\Theta(x+\beta t,t)\hfill
\\
&-\beta x-\frac{1}{2}\beta^2t;
\hfill
\\
\end{array}\hfill
&\hbox{boost}\\
\\
\begin{array}{c}
{x}^\star=x
\hfill
\\
{t}^\star=t,
\\
\end{array}\hfill
&\begin{array}{c}
R^\star(x,t)= R(x,t),\hfill
\\
\Theta^\star(x,t)=\Theta(x,t)-\eta.
\end{array}\hfill
&\hbox{phase shift}\\
\end{array}
\label{Galtr}
\end{equation}
change  the Lagrange density (\ref{BJ}) by a surface term,
and Noether's theorem yields the conserved quantities
(\ref{Galconst}).
The new conserved quantities (\ref{BJconst}) belong in turn
to the following
strange, non--linear action on space-time \cite{BJ}
\begin{equation}
\begin{array}{cc}
\begin{array}{c}
{x}^\star=
x+\alpha\Theta({x}^\star,{t}^\star)
\hfill
\\
{t}^\star=t+\frac{1}{2}\alpha
\big(x+{x}^\star)
\\
\end{array}\qquad
\hfill
&\hbox{``antiboost''}
\\
\cr
\begin{array}{c}
{x}^\star =x
\hfill
\\
t^\star =e^{\delta}t\hfill
\\
\end{array}
\hfill
&\hbox{time dilatation}
\\
\end{array}
\label{BJtransf}
\end{equation}

 ``Antiboosts'' are particularly interesting~:
 $x^\star $ and
${t}^\star $ are only defined implicitely,
and the action is ``field--\-depen\-dent''
in that its very definition  involves $\Theta$.

When implemented on the fields non--conventionally,
these transformations act as symmetries.
In detail, let us set
\begin{equation}
\begin{array}{cc}
\begin{array}{c}
R^\star (x,t)=
\displaystyle{\frac{ R({x}^\star, {t}^\star)}{J^\star}},\hfill
\\
\Theta^\star(x,t)=\Theta({x}^\star,{t}^\star);
\qquad\hfill
\\
\end{array}\hfill
&\hbox{``antiboost''}\hfill
\\
\cr
\begin{array}{c}
R^\star(x,t)=
e^{-\delta} R({x}^\star,{t}^\star),
\qquad\hfill
\\
\Theta^\star(x,t)=
e^{\delta}\Theta({x}^\star,{t}^\star);\hfill
\\
\end{array}\hfill
&\hbox{time dilatation}\hfill
\label{BJimp}
\end{array}
\end{equation}
where
$J^\star=
\big(1-\alpha\partial_{{x}^\star}\Theta({x}^\star,{t}^\star)
-\frac{1}{2}
\alpha^2\partial_{{t}^\star}\Theta({x}^\star,
{t}^\star)\big)^{-1}$
is the Jacobian of the space-time transformation
(\ref{BJtransf}). Then
the Lagrangian (\ref{BJ}) changes by a surface term and
the conserved quantities (\ref{BJconst}) are
recovered by Noether's theorem.

So far we merely reviewed the results from Ref. \cite{BJ}.
Now we point
out that, in the free case $V=0$, the system
described by the Lagrangian ${\cal L}_{0}$ in (\ref{BJ})
has even more symmetries.
Let us first remember that
the ``non-relativistic conformal transformations''
\begin{equation}
\begin{array}{cc}
\begin{array}{c}
{x}^\star =e^{\lambda/2}x,\hfill
\\
{t}^\star =e^{\lambda}t;\hfill
\\
\end{array}
\hfill
&\hbox{non-relat. dilatations}\hfill
\\
\cr
\begin{array}{c}
{x}^\star =\displaystyle\frac{x}{1-\kappa t},
\hfill
\\
{t}^\star =\displaystyle\frac{t}{1-\kappa t};\qquad
\hfill
\end{array}
&\hbox{expansions}\hfill
\\
\end{array}
\label{nrctransf}
\end{equation}
are symmetries for
the free Schr\"odinger equation \cite{JHN}.
The transformations in (\ref{nrctransf}) span
with the time translation $t^\star=t+\epsilon$
an ${\rm SL}(2,\RR)$ group;
added to the Galilei transformations (\ref{Galtr}), the
Schr\"odinger group is obtained.

Implementing (\ref{nrctransf}) on $R$ and $\Theta$ as
\begin{equation}
\begin{array}{cc}
\begin{array}{c}
R^\star(x,t)=
\displaystyle{e^{\lambda/2}} R({x}^\star,{t}^\star),
\hfill
\\
\Theta^\star(x,t)=
\Theta({x}^\star,{t}^\star);\hfill
\\
\end{array}\hfill
&\hbox{non-relat. dilatation}\hfill
\\
\\
\begin{array}{c}
R^\star(x,t)=
\displaystyle{\frac{1}{1-\kappa t}}\, R({x}^\star,{t}^\star),
\hfill
\\
\Theta^\star(x,t)=
 \Theta({x}^\star,{t}^\star)-
\displaystyle\frac{\kappa x^2}{2(1-\kappa t)};\hfill
\\
\end{array}\hfill
&\hbox{expansion}\hfill
\end{array}
\label{nrctrimp}
\end{equation}
 the free action is left invariant.
Thus, the transformations in (\ref{nrctransf})
act as symmetries also in our case.
The associated conserved quantities read
\begin{equation}
\begin{array}{cc}
\Delta=\displaystyle\int dx\,
\big(t{\cal H}-\frac{1}{2}x{\cal P}\big),\hfill
&\hbox{non-relativistic dilatation}\hfill
\\
\cr
K=-t^2H+2t\Delta
+\frac{1}{2}\displaystyle\int dx\,
x^2 R;\qquad\hfill
&\hbox{expansion}\hfill
\hfill
\label{newnrconst}
\end{array}
\end{equation}

Remarkably, the ``relativistic''
dynamical Poincar\'e symmetry can also
be conformally extended. Using the equations of motion
(\ref{eqmotion}), a lengthy but straightforward calculation shows
that, for $V=0$,
\begin{equation}
\begin{array}{c}
{\rm C}_{1}=\displaystyle\int dx\,\Big(
\frac{x^2}{2}{\cal H}-x\Theta{\cal P}+\Theta^2 R
\Big),\hfill
\\
{\rm C}_{2}=\displaystyle\int dx\Big(
xt{\cal H}-\big(\frac{x^2}{2}+t\Theta\big){\cal P}
+x\Theta R\Big)\;\quad
\hfill
\label{newrelconst}
\end{array}
\end{equation}
are also conserved, $\frac{d{\rm C}_{i}}{dt}=0$.
A shorter proof can be obtained by calculating
the energy--momentum tensor for (\ref{BJ}),
\begin{equation}
\begin{array}{c}
T_{tt}=
\displaystyle{\frac{ R}{2}}\big(\partial_{x}\Theta\big)^2
+cR^\omega,
\hfill
\ccr
T_{xt}=- R\partial_{x}\Theta\,\partial_{t}\Theta,\hfill
\ccr
T_{tx}= R\partial_{x}\Theta,\hfill
\ccr
T_{xx}= R\big(\partial_{x}\Theta\big)^2+(\omega-1)c R^\omega.
\hfill
\\
\end{array}
\label{BJem}
\end{equation}
(Let us note that in the
non--relativistic context the
usual index gymnastics is meaningless, since space-time does
not carry
a metric. We agree therefore that the
energy--momentum tensor $T_{\alpha\beta}$ is a covariant
two tensor
carrying lower indices, while $T^{\alpha\beta}$ is not defined).

The tensor $T_{\alpha\beta}$ is neither symmetric
nor traceless. It is nevertheless conserved,
$
\p_{\alpha}T_{\alpha\beta}=0
$
for all $\beta=t, x$. Let us rewrite the quantities
${\rm C}_{1}$ and ${\rm C}_{2}$ as the integrals of
\begin{equation}
\begin{array}{cc}
\displaystyle{\frac{x^2}{2}}T_{tt}^0-x\Theta T_{tx}^0+\Theta^2
R\qquad\hfill
&{\rm C}_{1},\hfill
\ccr
xtT_{tt}^0-\big(\displaystyle{\frac{x^2}{2}}+t\Theta\big)
T_{tx}^0
+x\Theta R\qquad\hfill
&{\rm C}_{2},\hfill
\\
\end{array}
\label{Ttransc}
\end{equation}
where $T_{\alpha\beta}^0$ denotes the free ($V=0$) energy-momentum
tensor.
Then the conservation of the quantities (\ref{newrelconst}) is
obtained by
 deriving this expression w. r. t. time and using the continuity
equation $\p_{\alpha}T_{\alpha\beta}^0=0.$

The Poisson brackets of our conserved quantities
(listed in Appendix A)
yield a closed, finite--dimensional algebra.
In the next Section we prove that this is in fact
the $\o(3,2)$ conformal algebra.

For the membrane potential $V=c/R$,
the conformal symmetries $\Delta$, $K$,
${\rm C}_{1}$ and ${\rm C}_{2}$
are broken, and only the Poincar\'e symmetry survives.
Changing the question, we can also ask for
 what potentials do we have the same symmetries as in the
 free case.
Now the Lagrangian (\ref{BJ}) is dilation
and indeed Schr\"odinger invariant only for
\begin{equation}
V=cR^3.
\label{confinvpot}
\end{equation}
This comes from the scaling properties of the Lagrange density,
and can also be seen of by looking at the
energy-momentum tensor (\ref{BJem})~:
the trace condition
\begin{equation}
T_{xx}=2T_{tt},
\label{tracecond}
\end{equation}
which is the signal for a Schr\"odinger symmetry \cite{JHN},
only holds
for $\omega=3$.
On the other hand, the potential $c R^\omega$ yields an
``antiboost--invariant''
expression only  for $\omega=-1$ so that he
Poincar\'e symmetry only allows the ``membrane potential''
(\ref{membpot}),
$
V(R)={c}/{R}.
$
Therefore, the full $\o(3,2)$ conformal symmetry only arises in
the free case.

\goodbreak
\section{A ``Kaluza-Klein'' framework}

In order to explain  the origin of the symmetries of the model,
let us start with Duval's unpublished observation \cite{Duval}.
Let us enlarge space-time by adding
a new, ``phase-like'' coordinate  $s$ i.e.,
consider the ``extended space''
\begin{equation}
M=\left\{\left(
\begin{array}{c}
x\\
t\\
s
\end{array}
\right)
\right\}.
\end{equation}

Let us lift the space--time transformations
to $M$ by adding a transformation rule for $s$
 inspired from the
rule the phase changes in Eq. (\ref{BJimp}).
Thus, let us formally replace
  the field $\Theta^\star(x,t)$ by the coordinate $-s$,
\begin{equation}
\Theta^\star(x, t)\to-s.
\label{thetarule}
\end{equation}

When applied to an ``antiboost'', for example,
 we get
the {\it linear} action on extended space\footnote{
Our notations are as follows.
$\mu, \nu, \dots = x, t, s$
are  indices on the extended space $M$, and
$\alpha, \beta \dots = x, t$
are indices on ordinary space--time, $Q$. The
transformed coordinates are denoted by ``tilde''
($\widetilde{\{\,\cdot\,\}}$) on  $M$,
while they are denoted by ``star'', ($\{\,\cdot\,\}^\star$),
on $Q$.
The fields on $M$ are denoted by lower--case letters
(e. g., $\rho, \, \theta$),
while the fields on $Q$ are
 denoted by upper--case letters like $R,\, \Theta$.},
$\left(\begin{array}{c}
x\\
t\\
s
\end{array}\right)\to
\left(\begin{array}{c}
\widetilde{x}\\
\widetilde{t}\\
\widetilde{s}
\end{array}\right)$,
\begin{equation}
\begin{array}{cc}
G~:\qquad\hfill
&\begin{array}{c}
\widetilde{x}=
x-\alpha s,
\hfill
\\
\widetilde{t}=t+\alpha x-\frac{1}{2}\alpha^2s,
\hfill
\\
\widetilde{s}=s.\hfill
\\
\end{array}
\hfill
\end{array}
\label{BJBtransf}
\end{equation}

On the other hand, lifting the action of galilean boosts to our
extended space-time by applying the same rules, we get
\begin{equation}
\begin{array}{cc}
B~:\qquad\hfill
\begin{array}{c}
\widetilde{x}=x+\beta t,\hfill
\\
\widetilde{t}=t,\hfill
\\
\widetilde{s}=s-\beta x-\frac{1}{2}\beta^2t.\hfill
\\
\end{array}
\hfill
\end{array}
\label{GalBtr}
\end{equation}

The action of the mysterious ``antiboost'' becomes hence
analogous to that of galilean boost, the only difference being
that ordinary time, $t$, and the new, phase-like coordinate, $s$,
have to be interchanged,
\begin{equation}
t\longleftrightarrow s.
\label{tsinterchange}
\end{equation}

When interchanging $t$ and $s$,
the dilations of time alone in Eq. (\ref{BJtransf})
lifted to extended space by the same rule as above, remain
dilations of time alone but with the inverse parameter~:
$\delta\to-\delta$,
\begin{equation}
\begin{array}{cccc}
D~:\;\hfill
&\begin{array}{c}
\widetilde{x}=x,
\hfill
\\
\widetilde{t}=e^{\delta}t,\hfill
\\
\widetilde{s}=e^{-\delta}s
\end{array}
&\Longrightarrow
&\begin{array}{c}
\widetilde{x}=x,
\hfill
\\
\widetilde{t}=e^{-\delta}t,\hfill
\\
\widetilde{s}=e^{\delta}s.
\end{array}
\end{array}
\label{BJBtransf2}
\end{equation}

This same rule changes a time translation with parameter
$\epsilon=-\eta$
into a the phase translation,
\begin{equation}
\begin{array}{ccc}
\hbox{time translation}\qquad\hfill
&
&\hbox{phase translation}
\\
\widetilde{x}=x\hfill
&
&\widetilde{x}=x\hfill
\\
\widetilde{t}=
t+\epsilon\hfill
&\Longrightarrow\qquad\hfill
&\widetilde{t}=t\hfill
\\
\widetilde{s}=s
\qquad\hfill
&
&\widetilde{s}=s-\eta\hfill
\end{array}
\label{tvtr}
\end{equation}
\vskip-2mm
i.e.,
\begin{eqnarray*}
\vbox{\halign{#\qquad\qquad &#\qquad\qquad &#
\cr
energy\hfill
&$\Longrightarrow$\hfill
&particle number
\cr
}
}
\end{eqnarray*}

Our trick of adding an extra coordinate $s$ allowed us so far
to reconstruct the Poincar\'e group from the extended
Galilei group
by the ``interchange rule'' (\ref{tsinterchange}).
The conformal extensions can be similarly investigated.
Non--relativistic dilations act as
\begin{equation}
\begin{array}{ccc}
\Delta:
\left(
\begin{array}{c}
x\\
t\\
s\\
\end{array}\right)
&\to
\hfill
&\left(\begin{array}{c}
e^{\lambda/2}x,\hfill
\\
e^{\lambda}t;\hfill
\\
s
\end{array}\right).\hfill
\end{array}
\end{equation}

Let us observe that relativistic dilations, i. e.,
uniform dilations of all coordinates,
$
d:\left(
\begin{array}{c}
x
\\
t
\\
s
\end{array}\right)
 \to
\left(\begin{array}{c}
e^{\delta}x
\\
e^{\delta} t\\
e^{\delta}s
\\
\end{array}\right)
$
 also belong to our algebra, since they correspond to a
non-relativistic dilation ($\Delta$) with parameter
$2\delta$,
followed by
a dilation of time alone ($D$) with parameter $-\delta$,
$
d=D_{-\delta}
\smallcirc\Delta_{2\delta}.
$
Then the  $t\leftrightarrow s$ counterpart of a
non--relativistic dilation is
a uniform dilation followed by a dilation of time alone,
\begin{equation}
\Delta_{\lambda}\to
D_{-\lambda/2}
\smallcirc d_{\lambda/2}.
\end{equation}
The $s\leftrightarrow t$ counterpart of  non--relativistic
expansions (\ref{nrctransf})--(\ref{nrctrimp}) with parameter
$\kappa=-\epsilon_{1}$ is in turn a new transformation
we denote by ${\rm C}_{1}$,
\begin{equation}
\begin{array}{ccc}
\hbox{expansions}\qquad\hfill
&
&{\rm C}_{1}
\\
\cr
\widetilde{x}=
\displaystyle\frac{x}{1-\kappa t}\hfill
&
&\widetilde{x}=
\displaystyle\frac{x}{1+\epsilon_{1}s}\hfill
\\
\widetilde{t}=
\displaystyle\frac{t}{1-\kappa t}\hfill
&\Longrightarrow\quad\hfill
&\widetilde{t}=t+
\displaystyle\frac{\epsilon_{1}x^2}{2(1+\epsilon_{1}s)}
\hfill
\\
\widetilde{s}=s-
\displaystyle\frac{\kappa x^2}{2(1-\kappa t)}
\qquad\hfill
&
&\widetilde{s}=
\displaystyle\frac{s}{1+\epsilon_{1} s}\hfill
\\
\end{array}.
\label{expC1}
\end{equation}

The infinitesimal version of the new transformation is
\begin{equation}
X_{8}=\frac{x^2}{2}\p_{t}-xs\p_{x}-s^2\p_{s}.
\label{X8}
\end{equation}

Calculating the Lie brackets of (\ref{X8}) with the other
infinitesimal transformations, we get one more
vectorfield. In fact, the bracket of (\ref{X8}) with
the generator of
infinitesimal boosts, $t\p_{x}-x \p_{s}$, yields
\begin{equation}
X_{9}
=
xt\p_{t}+\big(\frac{x^2}{2}-ts\big)\p_{x}+xs\p_{s}.
\label{X9}
\end{equation}

Collecting our results, our symmetry generators read
\begin{equation}
\begin{array}{cccc}
X_{0}\hfill
&=
&\p_{t}\hfill
&\hbox{time translation}\hfill
\\
X_{1}\hfill
&=
&-\p_{x}\hfill
&\hbox{space translation}\hfill
\\
X_{2}\hfill
&=
&-\p_{s}\hfill
&\hbox{vertical translation}\hfill
\\
X_{3}\hfill
&=
&t\p_{x}-x\p_{s}\hfill
&\hbox{galilean boost}\hfill
\\
X_{4}\hfill
&=
&t\p_{t}+\frac{x}{2}\p_{x}\hfill
&\hbox{non-relat. dilatation}\hfill
\\
X_{5}\hfill
&=
&t^2\p_{t}+xt\p_{x}-\frac{x^2}{2}\p_{s}\hfill
&\hbox{expansion}\hfill
\\
X_{6}\hfill
&=
&t\p_{t}-s\p_{s}\hfill
&\hbox{time dilation}\hfill
\\
X_{7}\hfill
&=
&x\p_{t}-s\p_{x}\hfill
&\hbox{``antiboost''}\hfill
\\
X_{8}\hfill
&=
&\frac{x^2}{2}\p_{t}-xs\p_{x}-s^2\p_{s}\hfill
&{{\rm C}_{1}}\hfill
\\
X_{9}\hfill
&=
&xt\p_{t}+\big(\frac{x^2}{2}-ts\big)\p_{x}+xs\p_{s}\qquad
\hfill
&{\rm C}_{2}\hfill
\\
\end{array}
\label{Bgenerators}
\end{equation}

The Lie brackets of these vector fields are seen to satisfy the
same algebra as the conserved quantities  in (\ref{Poissonbrackets}).
The vectorfields $X_8$ and
$X_{9}$ will be shown below in particular to generate
the two additional conserved quantites ${\rm C}_{1}$ and
${\rm C}_2$ in Eq.
(\ref{newrelconst}).
Note that the algebra (\ref{Bgenerators}) is manifestly invariant
w. r. t. the interchange $t\longleftrightarrow s$.
The vector field $X_{9}$ is itself invariant;
this is the reason why we could not find it by the
``interchange--trick''.

The extended manifold $M$ above has
already been met before.
 In their ``Kaluza-Klein-type'' framework
for non-relativistic physics in $d+1$ dimension,
Duval et al. \cite{barg}  indeed consider
 a $(d+1,1)$--dimensional Lorentz manifold
$\big(M, g_{\mu\nu}\big)$, endowed with a
covariantly constant lightlike ``vertical'' vector $\xi=(\xi^\mu)$
they call ``Bargmann space''.
The quotient of $M$ by the flow of $\xi$ is a non-relativistic
space-time denoted by $Q$.
In the application we have in mind, $M$ is simply
$3$-dimensional Minkowski space, with
the usual coordinates $x_0, x, y$ and metric
$-(dx^{0})^2+dx^2+dy^2$.
Introducing the light-cone coordinates
\begin{equation}
t=\frac{1}{\sqrt{2}}\big(y-x^{0}\big),
\qquad
s=\frac{1}{\sqrt{2}}\big(y+x^{0}\big),
\end{equation}
the Minkowskian metric reads
$dx^2+2dtds$. Then $\xi=\partial_s$
 is indeed lightlike and covariantly constant.

All [infinitesimal] conformal transformations of Minkowski space
form the conformal algebra $\o(3,2)$.
Now, as shown in Appendix A, the  $X_{i}$
found above provide just another basis
of this same algebra.

\goodbreak
\section{Conformal geometry}


The action of the orthogonal group $\O(3,2)$ on
$3$-dimensional Minkowski space is the best described as follows.
Consider the natural action of $\O(3,2)$ on $\RR^{3,2}$
by matrix multiplication. A vector in
$\RR^{3,2}$ can be written as
\begin{equation}
Y=\pmatrix{y\cr a\cr b\cr},
\where
y=\pmatrix{x\cr t\cr s}\in\RR^{2,1},
\;
a,\, b\in\RR.
\end{equation}

The vector space $\RR^{3,2}$ carries the quadratic form
$
{\bar Y}Y={\bar y}y+2ab,
$
where ${\bar y}y$ means
$
{\bar y}y=x^2+2ts,
$
so that $\bar{Y}$ is represented by the row-vector
$(\bar{y}, b, a)$ where
$\bar{y}=(x, s, t)$.
$(2+1)$-dimensional Minkowski space,
$M=\RR^{2,1}$, can be mapped into the isotropic
cone (quadric) ${\cal Q}$ in $\RR^{3,2}$, as
\begin{equation}
y\mapsto
\pmatrix{y\cr1\cr-\2{\bar y}y\cr}.
\end{equation}

Projecting onto the real projective space $P{\cal Q}$, we
identify $M$ with
those generators in the null-cone in $\RR^{3,2}$.
The manifold $P{\cal Q}$ is invariant with respect to the action
of $\O(3,2)$.

Let us first consider  infinitesimal actions.
 An $\o(3,2)$ matrix can be
written as
\begin{equation}
\pmatrix{
\Lambda &V&W\cr
-{\bar W}&-\lambda &0\cr
-{\bar V}&0 &\lambda\cr
},
\qquad
\Lambda\in\, \o(2,1),\,V,\,W\in\RR^{2,1},
\;\lambda\in\RR.
\label{o32matrix}
\end{equation}

The matrix action of $\o(3,2)$ on $\RR^{3,2}$ yields
the action on Bargmann space
\begin{equation}
\Lambda y+V-\2W{\bar y}y+({\bar W}y+\lambda)y.
\label{confalgebraaction}
\end{equation}

In particular, $V$ represents infinitesimal translations.
Observe now that the
covariantly constant null vector $\xi$ is also the
generator of vertical translations,
\begin{equation}
{\hat\xi}=\pmatrix{
0&\xi&0\cr
0&0&0\cr
-\bar{\xi}&0&0\cr
}.
\label{ximatrix}
\end{equation}

The Schr\"odinger algebra is identified as
 those vectorfields which commute with the ``vertical vector'',
\begin{equation}
[Z,{\hat\xi}]=0.
\end{equation}
 This yields the constraints
$\Lambda\xi=-\lambda\xi$
and
$W=\kappa\xi$,
$\kappa\in\RR$. It follows that
\begin{equation}
Z=\pmatrix{
0
&\beta &0&\gamma&0
\cr
0&\lambda &0&\tau &0
\cr
-{{\beta}}&0&-\lambda&\eta&\kappa\cr
0&-\kappa&0&-\lambda&0\cr
-{\gamma}&-\eta&-\tau&0&\lambda\cr
},
\qquad
\beta,\gamma, \lambda,\kappa,\eta\in\RR.
\end{equation}

This is the {\it extended Schr\"odinger algebra}, with
\vskip1mm

$\bullet$ ${\beta}$ representing Galilei boosts,

$\bullet$ ${\bf \gamma}$
space translations,

$\bullet$ $\tau$ time translations,

$\bullet$ $\lambda$ non-relativistic dilatations,

$\bullet$ $\kappa$ expansions,

$\bullet$ $\eta$ translations in the vertical direction.

\vskip1mm
Using (\ref{confalgebraaction}),
we recover the infinitesimal action of the (extended)
Schr\"o\-din\-ger
algebra on $M$ \cite{barg}.
Note that the {\it relativistic}
dilation invariance [with all directions dilated by the same factor], is
{\it broken} by the reduction:
only doubly time-dilated combinations project to Bargmann space.

Those  parameterized by
$\beta, \gamma, \tau,\eta\in\RR$ are isometries
and are recognized as the generators of the {\it extended Galilei}
 (or Bargmann)  group.

\goodbreak
Let us now identify the unusual generators.
 ``Antiboosts'' and dilatations of time alone belong
to the upper--left $\o(2,1)$ corner $\Lambda$ of
the $\o(3,2)$ matrix,
(\ref{o32matrix})
\begin{equation}
\begin{array}{cc}
\Lambda=\qquad\hfill
&\left\{\begin{array}{cc}
\left(\begin{array}{ccc}
0&0&-\alpha
\\
\alpha&0&0
\\
0&0&0
\end{array}\right)
\qquad\hfill
&\hbox{antiboost}\hfill
\\\cr
\left(\begin{array}{ccc}
0&0&0
\\
0&d&0
\\
0&0&-d
\end{array}\right)
\qquad\hfill
&\hbox{dilation of time alone}\hfill
\\
\end{array}\right.
\end{array}
\end{equation}
Augmented with the extended Galilei
algebra, the Poincar\'e algebra is obtained.

In the same spirit, the two remaining (relativistic)
conformal transformations ${\rm C}_{1}$ and ${\rm C}_{2}$
correspond to chosing
$
W_{1}=\left(\begin{array}{c}
0\\
1\\
0\\
\end{array}\right)
$
and
$
W_{2}=\left(\begin{array}{c}
1\\
0\\
0\\
\end{array}\right),
$
respectively.
The generated group, found by exponentiating, is the conformal
group
${\rm SO}(3,2)$.

The Schr\"odinger group is recovered as those
 transformations which commute with the
1-parameter subgroup generated by ${\hat\xi}$.
It acts on  $M$ according to
in the standard way which plainly
 project to ``ordinary'' space-time and span there
the (non--extended) Schr\"o\-din\-ger group consistently with
(\ref{Galtr}) and (\ref{nrctransf}).
The  transformations which do {\it not} preserve $\xi$ are
\begin{equation}
\begin{array}{cc}
\begin{array}{ccc}
\widetilde{x}&=\hfill &x\\
\widetilde{t}&=\hfill &e^{\delta}t
\\
\widetilde{s}&=\hfill &e^{-\delta}s
\end{array}\hfill
&\hbox{time dilation}\hfill
\\
\cr
\begin{array}{ccc}
\widetilde{x}&=\hfill &x-\alpha s
\\
\widetilde{t}&=\hfill &t+\alpha x-\frac{1}{2}\alpha^2s
\\
\widetilde{s}&=\hfill &s
\\
\end{array}\hfill
&\hbox{``antiboost''}\hfill
\\
\cr
\begin{array}{ccc}
\widetilde{x}&=\hfill &
\displaystyle\frac{x}{1+\epsilon_{1}s}
\\
\widetilde{t}&=\hfill &t+
\2\displaystyle\frac{\epsilon_{1}x^2}{1+\epsilon_{1}s}
\\
\widetilde{s}&=\hfill &
\displaystyle\frac{s}{1+\epsilon_{1}s}
\\
\end{array}
\hfill
&{\rm C}_{1}
\\
\cr
\begin{array}{ccc}
\widetilde{x}&=\hfill &
\displaystyle{
\frac{x-\epsilon_{2}\big(\2x^2+ts\big)}
{(1-\2\epsilon_{2}x)^2+\2\epsilon_{2}^2ts}}
\\
\widetilde{t}&=\hfill &
\displaystyle{
\frac{t}{(1-\2\epsilon_{2}x)^2+\2\epsilon_{2}^2ts}}
\\
\widetilde{s}&=\hfill &
\displaystyle{
\frac{s}{(1-\2\epsilon_{2}x)^2+\2\epsilon_{2}^2ts}}
\\
\end{array}
\qquad\hfill
&{\rm C}_{2}
\\
\end{array}
\label{Cgroupaction}
\end{equation}

Our transformations are indeed conformal since they satisfy
$
f^*g_{\mu\nu}=\Omega^2g_{\mu\nu}
$
see Appendix A.
Note that the interchange
$s\leftrightarrow t$ is also an isometry and
carries the group {\rm SO}$(3,2)$ into another component of the
conformal group ${\rm O}(3,2)$.

\goodbreak
\section{Projecting to ordinary space--time}

As we said already, the quotient $Q$ is $1+1$-dimensional
``ordinary'' spacetime,
labeled by $x$ (position) and $t$ (time).
The projection
$M\to Q$ means simply ``forgetting'' the vertical coordinate
$s$~:
$\left(\begin{array}{c}
x\\ t\\ s
\end{array}\right)
\to
\left(\begin{array}{c}
x\\ t
\end{array}\right).
$

Next, we wish to relate the fields on extended and on
ordinary space, respectively.
Let us recall how this is done usually \cite{barg}.
Let $\psi$ denote a complex field
 on $M$. Then, requiring the field to be equivariant,
\begin{equation}
\xi^\mu\p_{\mu}\psi=i\psi,
\label{oldequiv}
\end{equation}
allows us to reduce $\psi$
 from Bargmann space to one  on ordinary space-time as
$\Psi(x,t)=e^{-is}\psi(x,t,s)
$
\cite{barg}. Writing $\psi=\rho^{1/2}e^{i\theta}$,
(\ref{oldequiv}) reads
$
\xi^\mu\partial_\mu\rho=0
$
and
$
\xi^\mu\partial_\mu\theta=1.
$
In light--cone coordinates of Minkowski case in particular,
 these conditions imply that
\begin{equation}
R(x,t)=\rho(x,t,s),
\qquad
\Theta(x,t)=\theta(x,t,s)-s.
\label{modequiv}
\end{equation}
are well--defined fields on $Q$. These formul{\ae}
(also referred to as  equivariance) allows us to relate
equivariant
fields on extended space to fields on ordinary space.

Let us now consider a diffeomorphism
 \begin{eqnarray}
f(x,t,s)\equiv
\left(\begin{array}{c}
\widetilde{x}
\\
\widetilde{t}
\\
\widetilde{s}
\end{array}\right)
=
\left(\begin{array}{c}
g(x , t, s)
\\
h(x , t, s)
\\
k(x, t, s)
\end{array}\right)
\label{difofM}
\end{eqnarray}
of $M$. How can we
project this to ordinary space--time~?
In the particular case when the mapping preserves $\xi$,
the entire fibre goes into the same fibre and the result
projects to a well--defined diffeomorphism of ordinary
space-time.
In fact,
$
\widetilde{x}=g(x,t)
$,
$
\widetilde{t}=h(x,t)
$,
$\widetilde{s}\equiv k(x,t,s)=s+K(x,t)
$,
so that we can define the projected map
$F(x,t)=\left(\begin{array}{c}
x^\star
\\
t^\star\end{array}\right)
$
by setting
$
\begin{array}{c}
x^\star=\widetilde{x}=g(x,t)
$
and
$
t^\star=\widetilde{t}=h(x,t).
\end{array}
$
As a bonus, we also get
the usual transformation
rule of the phase (consistent with the equivariance),
$
\begin{array}{c}
\Theta^\star(x,t)=\Theta(x^\star,t^\star)+K(x,t).
\end{array}
$

If, however, $f$ does not preserve the fibres, this construction
does
not work since the coordinates $\widetilde{x}$ and
$\widetilde{t}$
now depend on $s$. Hence the need of generalizing the
construction
based on equivariance. Forgetting momentarily about
$\rho$, we only
consider the phase, $\theta$.
Our clue is to observe that if $\theta$ is equivariant,
(\ref{modequiv}),
then $s=-\Theta(x,t)$ is solution of the equation
$\theta(x,t, s)=0$, i. e.,
\begin{equation}
\theta\big(x,t,-\Theta(x,t)\big)=0.
\label{newequiv}
\end{equation}

This condition is, however, meaningful without any assumption of
equivariance and associates implicitly a function
$\Theta(x,t)$ to each
$x$, $t$ {\it and} field $\theta$. Conversely, to any
$x$, $t$  and  $\Theta(x,t)$ Eq. (\ref{modequiv}) associates an
(equivariant) field $\theta(x,t,s)$ on $M$.

Let us recall that
our extended ``Bargmann'' space $M$ is  a fibre bundle over
ordinary space-time $Q$, with fibre $\RR$. Then
$\Theta$ corresponds to a section $Q\to M$ of this bundle.
Condition (\ref{newequiv}) requires the existence of a section
$\Theta(x,t)$ along which the phase field $\theta$ vanishes.

A diffeomorphism $f$
of $M$ acts naturally on $\theta$, namely as
\begin{equation}
\widetilde{\theta}=f^\star\theta.
\label{actionontheta}
\end{equation}
We can define therefore $\Theta^\star$ as the solution
of the equation
\begin{equation}
\widetilde\theta(x,t,-\Theta^\star(x,t))=0.
\label{projcond}
\end{equation}
This implicit equation (assumed to admit a unique
solution) associates a $\Theta^\star$ to $x$, $t$ and
$\theta$.
Let us stress that
$\widetilde{\theta}$ is not in general equivariant even
if $\theta$ is equivariant, unless $f$ preserves the fibres.

Thus,
starting with $\Theta(x,t)$ we lift it first to $M$ as an
equivariant field $\theta(x,t,s)=\Theta(x,t)+s$ on $M$; to
which a
well--defined $\Theta^\star$ (function of
$x$, $t$ and $\Theta$) is associated by (\ref{projcond}).
Having defined
$\Theta^\star$,
the diffeomorphism $f$ of $M$
can be projected to $Q$ in a ``field--dependent way'' by
restricting
$\widetilde{f}$ to the section $s=-\Theta^*$. In
coordinates,
$F(x,t)=
\left(\begin{array}{c}
x^\star
\\
t^\star
\\
\end{array}\right)
$, where
\begin{equation}
\begin{array}{c}
x^\star=g\big(x,t,-\Theta^\star(x,t)\big),
\\
t^\star=h\big(x,t,-\Theta^\star(x,t)\big),
\\
\Theta(x^\star,t^\star)=
-k\big(x,t,-\Theta^\star(x,t)\big).\hfill
\end{array}
\label{actiononQ}
\end{equation}

The last line here requires to express
$\Theta^\star$ by inverting the function $k$ and
reinserting the result
into the two first lines.
It also implements the transformation
on the ``phase'', $\Theta$.
Let us stress that these formul{\ae} are implicit~:
$x^\star$ and $t^\star$ can not be defined without defining
${\Theta}^\star $, which itself involves $x^\star$ and $t^\star$.

In the equivariant case, the procedure is
plainly consistent with the previous formulae.

In the non-fiber-preserving case
 it yields the ``field--dependent
diffeomorphisms'' considered by Bazeia and Jackiw \cite{BJ}.
For ``antiboosts'', for example, we get from
(\ref{BJBtransf})
\begin{equation}
\begin{array}{cc}
&{x}^\star=
x+\alpha{\Theta}^\star (x, t),
\hfill
\\
&{t}^\star=t+\alpha x+\frac{1}{2}\alpha^2
{\Theta}^\star (x, t)
\\
&{\Theta}^\star(x, t)=
\Theta(x^\star, t^\star),
\hfill
\end{array}
\label{ABdef}
\end{equation}
which is equivalent to the definition (\ref{BJtransf}).
Time dilations work similarly.
The formulae valid for  the
two relativistic conformal transformations,
${\rm C}_{1}$ and ${\rm C}_{2}$ above,  is presented in
 Appendix B (\ref{rctransf}).
The formula for ${\rm C}_{1}$ is consistent with (\ref{expC1});
that for ${\rm C}_{2}$ is a new result.

So far, we only studied how to act on $\Theta$~:
 (\ref{actiononQ}) only involves the phase but not
the density.
Turning to this problem, let us {\it posit}
\begin{equation}
R(x,t)=
\rho\big(x,t,-\Theta(x,t)\big)
\p_{s}\theta\big(x,t,-\Theta(x,t)\big),
\label{Rdef}
\end{equation}
where $\Theta$ is defined by (\ref{newequiv}).
$R(x,t)$ is a well--defined function of $x$ and $t$.
Let us insist that (\ref {Rdef}) is again ``field--dependent'' in that it
also depends on $\theta$, except when
 $\theta$
is equivariant, when it reduces to (\ref{modequiv}).
Conversely, if $R(x,t)$ is any field on $Q$,
$\rho(x,t,s)=R(x,t)$ can
obviously be viewed as
(an equivariant) function on extended space.

Let us henceforth consider a conformal transformation $f$
of $M$ $f^\star g_{\mu\nu}=\Omega^2g_{\mu\nu}$ and let $\rho$
be a (possibly not equivariant) field on $M$.
$f$ acts naturally on $\rho$ as
\begin{equation}
\rho\to\widetilde{\rho}=\Omega f^*\rho.\hfill
\label{Brhoimpl}
\end{equation}
Hence
\begin{equation}
R^\star(x,t)=
\widetilde{\rho}\big(x,t,-\Theta^\star(x,t)\big)
\p_{s}\widetilde{\theta}\big(x,t,-\Theta^\star(x,t)\big).
\label{Rstardef}
\end{equation}

Using the definition (\ref{Rdef}) of $R$, this is also
written as
\begin{equation}
R^\star(x,t)=
\Omega(x,t,-\Theta^\star)\,
\frac{\p_{s}\widetilde{\theta}
\big(x,t,-{\Theta}^\star(x,t)\big)}
{\p_{\widetilde{s}}\theta\big(x^\star,t^\star,
-{\Theta}(x^\star,t^\star)\big)}
R({x}^\star,{t}^\star).
\label{Rimp}
\end{equation}
(If the field $\theta$ is equivariant,
the denominator is equal to $1$).
On the other hand, one can show in general that
\begin{equation}
\frac{\p_{s}\widetilde{\theta}
\big(x,t,-{\Theta}^\star(x,t)\big)}
{\p_{\widetilde{s}}\theta\big(x^\star,t^\star,
-{\Theta}(x^\star,t^\star)\big)}
=
\frac{\widetilde{J}\big(x,t,-{\Theta}^\star(x,t)\big)}
{J^\star(x,t)},
\end{equation}
where ${J}^*$ and
$\widetilde{J}$
are the Jacobians on ordinary and on the extended space
respectively,
\begin{equation}
\begin{array}{cc}
J^\star
={\rm det}\left(
\displaystyle{\frac{\p\big(x^\star\big)^\alpha}
{\p x^\beta}}
\right)\qquad
\hfill
&\widetilde{J}
={\rm det}\left(
\displaystyle{\frac{\p\widetilde{x}^\mu}
{\p x^\nu}}\right),
\end{array}
\label{jaconB}
\end{equation}
($\alpha, \beta=x, t$ and $\mu,\nu=x, t, s$).
Eq. (\ref{Rimp}) can therefore be rewritten as
\begin{equation}
R^\star(x,t)
=
\Omega\big(x,t,-{\Theta}^\star(x,t)\big)\times
\frac{\widetilde{J}\big(x,t,-{\Theta}^\star(x,t)\big)}
{J^\star(x,t)}\,R(x^\star,t^\star).
\label{imponR2}
\end{equation}

For a conformal transformation $\widetilde{J}=\pm \Omega^3$,
the
sign depending on the mapping being orientation--preserving
or not.

If the transformation $f$ preserves $\xi$, $\Omega$ is a
function
of  $t$--alone cf. (\ref{conffactor}) in Appendix A. Then
 $\widetilde{\theta}$ is
again equivariant and our formula reduces to the standard
expression
\begin{equation}
R^\star(x,t)
=
\Omega(t)\,
R(x^\star,t^\star),
\label{imponR3}
\end{equation}
cf. (\ref{nrctrimp}).
For an isometry, $\Omega=1$, so that
 (\ref{imponR2}) reduces to
\begin{equation}
R^\star(x,t)
=
\frac{R(x^\star,t^\star)}
{J^\star(x,t)}.
\end{equation}

For time dilations
and ``antiboosts'',
the formul{\ae} of Bazeia and Jackiw in \cite{BJ},
(our (\ref{BJconst})), are recovered.
For the relativistic conformal transformations
${\rm C}_{1}$ and ${\rm C}_{2}$,
we find some complicated expressions
(\ref{imppr}), presented in Appendix B.

Our formulae allow to implement any
isometry of $M$, not only those in
the connected component of the Poincar\'e
group. Let us consider, for example, the interchange
\begin{equation}
t\longleftrightarrow s,
\end{equation}
which is a non--fiber-preserving isometry. It acts on the
fields defined on $M$ in the natural way.
For fields on $Q$, we get the ``field-dependent action''
\begin{equation}
\begin{array}{c}
x^\star=x,
\ccr
t^\star=-\Theta^\star(x,t),
\ccr
\Theta\big(x,-\Theta^\star\big)+t=0,
\ccr
R^\star(x,t)=R\big(x,-\Theta^\star\big)\p_{t}
\Theta(x,-\Theta^\star)
=\displaystyle{
\frac{R\big(x,-\Theta^\star\big)}{\p_{t}\Theta^\star(x,t)}}.
\end{array}
\label{stimpl}
\end{equation}

This formula is so much implicit that we can not go farther unless
$\Theta$ is given explicitly. It is nevertheless a ``field--dependent
symmetry''.

The weak condition (\ref{projcond}) can hence accomodate the
$t\leftrightarrow s$ symmetry.
This is in sharp contrast with the equivariance condition
(\ref{oldequiv}) which manifestly breaks it.
Note also that our formul{\ae} for implementing
the conformal transformations on the fields are consistent
with the
interchange symmetry $t\leftrightarrow s$, followed by the rule of replacing
$s$ with $-\Theta^\star$.
When applied to a Schr\"odinger transformation, it yields
its non--$\xi$--preserving counterpart.

\section{Physics on extended space}

So far the ``Bargmann space'' $M$ was only used as a geometric
arena for linearizing the action of the conformal group.
Now we show how to lift the {\it physics} to $M$.
Generalising our
previous theory, let  $M$ be
$(d+1,1)$--dimensional Lorentz manifold
$\big(M, g_{\mu\nu}\big)$ endowed with a
covariantly constant lightlike  vector $\xi=(\xi^\mu)$.
Such a manifold admits a preferred coordinates
$\vec{x}, t, s$ in which the metric is
\begin{equation}
g_{ij}(\vec{x},t)dx^idx^j+2dt\big[ds+\vec{A}\cdot d\vec{x}\big]
-2U(\vec{x},t)dt^2,
\label{BRINK}
\end{equation}
where $g_{ij}$ is a metric on $d$--dimensional
``transverse space''
and $\vec{A}$ and $U$ are vector and scalar potentials,
respectively \cite{BRINK}, \cite{barg}.

\subsection{Field theory on extended space}

Let $\rho$ and $\theta$ be two real fields on $M$, and
let us consider the field theory described by the
action
\begin{equation}
\begin{array}{cc}
S=&S_{0}+S_{p}=\hfill
\ccr
&\displaystyle{
\int-\frac{1}{2}\left(\rho\nabla_\mu\theta\nabla^\mu\theta
\right)
\sqrt{-g}\,d^3x
-\int
V(\rho)\sqrt{-g}\,d^3x},
\hfill
\end{array}
\label{barlag}
\end{equation}
where $\nabla_{\mu}$ is the covariant derivative associated
with the
metric of $M$.
The Euler-Lagrange equations read
\begin{equation}
\nabla_\mu(\rho\nabla^\mu\theta)=0,
\qquad
\2\nabla_\mu\theta\nabla^\mu\theta=-\frac{dV}{d\rho}.
\label{bareq}
\end{equation}

When the fields are required to be  also equivariant,
 (\ref{modequiv}), then, for the projected variables
$\Theta$ and $\rho$, the equations of motion (\ref{bareq})
reduce to
those of Bazeia and Jackiw in Ref. \cite{BJ},
Eqn. (\ref{eqmotion}) above. (Working with a general
Bargmann space \cite{barg} would allow us to describe our
fluid system in
an external electromagnetic field).

Equivariance is a too strong condition, though.
For specific potentials, the weaker conditions (\ref{newequiv})
and
(\ref{Rimp}), i.e.
\begin{equation}
\begin{array}{c}
\theta\big(x,t,-\Theta(x,t)\big)=0,\hfill
\\
R(x,t)=
\rho\big(x,t,-\Theta(x,t)\big)
\p_{s}\theta\big(x,t,-\Theta(x,t)\big)\hfill
\end{array}
\label{weakc}
\end{equation}
may still work. Expressing $\p_{\alpha}\Theta$ by deriving
the defining relation (\ref{newequiv}) one finds using the
Euler--Lagrange equations (\ref{bareq}) that $R$ and $\Theta$
satisfy the Bazeia--Jackiw (\ref{eqmotion}) {\it provided}
$V(\rho)$ is the membrane potential $V(\rho)=c/\rho$. This
is hence
the only potential consistent with (\ref{newequiv}).

In sharp contrast with equivariance,
our new condition does not impose any restriction to the fields.
 Let us consider, for  example, the action
(\ref{barlag}) on $(2+1)$ dimensional Minkowski space with
$V=0$ and chose
\begin{equation}
\rho=\sqrt{R}
\qquad\hbox{and}\qquad
\theta=\sqrt{R}\,\sin(\Theta+s).
\end{equation}
The corresponding field on Bargmann space,
$\psi={R}^{1/4}e^{i\sqrt{R}\,\sin(\Theta+s)}$, is not equivariant.
This Ansatz satisfies however our conditions (\ref{weakc}),
as anticipated by
the notations. It projects to (\ref{BJ})
with its large symmetry.

\goodbreak
\subsection{Symmetries}

The Kaluza-Klein type framework is particularly
convenient for studying the symmetries.
Let us indeed consider a conformal diffeomorphism
$f(x,t,s)$ of the Bargmann metric.
It is easy to see, along the lines indicated in
Refs. \cite{barg}, that implementing $f$ on the fields as
\begin{equation}
\begin{array}{c}
\theta\to\widetilde{\theta}=f^*\theta,\hfill
\\
\rho\to\widetilde{\rho}=\Omega f^*\rho,\hfill
\end{array}
\label{impl}
\end{equation}
the  ``free'' action (\ref{barlag}) is left
invariant by all conformal transformations of $M$.
This is explained by the absence of any mass term in
(\ref{barlag}).
Equivalently, the transformed fields are seen to satisfy
the equations of motion
\begin{equation}
\nabla_\mu(\widetilde\rho\,\nabla^\mu\widetilde\theta)=0,
\qquad
\nabla_\mu\widetilde\theta\,\nabla^\mu\widetilde\theta=0.
\label{eqonB}
\end{equation}

It is worth noting that
unfolding to extended space converted the
up--to-surface--term invariant system (\ref{BJ}) into a
strictly invariant one.

\goodbreak
We can now derive once again the symmetries
starting from the extended space. Let us first consider the
free case.
Differentiating the defining relations
(\ref{projcond}) and (\ref{Rstardef})
we find, using the equations of motion (\ref {eqonB}) on $M$,
 that $R^\star$ and $\Theta^\star$
satisfy the free equations of motion in ordinary space,
\begin{eqnarray*}
\partial_{t}R^\star+\partial_{x}\big(
R^\star\partial_{x}\Theta^\star\big)=0,
\qquad
\partial_t\Theta^\star+\frac{1}{2}
\big(\partial_x\Theta^\star\big)^2=0.
\end{eqnarray*}

Alternatively, we check readily that
\begin{eqnarray*}
dxdtR^\star(x,t)\Big[\p_{t}\Theta^\star(x,t)
+\frac{1}{2}\big(\p_{x}\Theta^\star(x,t)\big)^2\Big]
=\hfill
\\
dx^\star dt^\star R(x^\star, t^\star)
\Big[\p_{t^\star}\Theta(x^\star, t^\star)
+\frac{1}{2}\big(\p_{x^\star}\Theta(x^\star,t^\star)\big)^2
\Big].
\hfill
\end{eqnarray*}
 The free action (\ref{BJ}) is hence invariant~:
 each conformal transformation of extended space
projects to a symmetry of the free system.
Restoring the potential term,
the scaling properties imply again that
conformal symmetry on $M$ only allows
$
V=c\rho^3,
$
cf (\ref{confinvpot}).
This potential is, however, inconsistent with the generalized
condition (\ref{newequiv}) unless $c=0$. Then we have the choice~:
if we keep $V=c\rho^3$ and use the
usual equivariance (\ref{oldequiv}), then the non--fiber preserving
part is broken and we are left with a Schr\"odinger symmetry.
If we choose $V=c/\rho$ the conformal symmetry is broken to its
Poincar\'e subgroup from the outset; this survives, however, the
reduction based on the generalised condition (\ref{newequiv}).
In particular, the interchange
$t\leftrightarrow s$,
 implemented as in (\ref{impl}) on $\theta$ and $\rho$
(or on $\Theta$ and $R$ as in (\ref{stimpl})) is a symmetry.

\goodbreak
\subsection{Conserved quantities}

On Bargmann space, we have a relativistic theory.
Defining the energy--momentum tensor
 as the variational derivative of
the action w. r. t. the metric,
$
\T_{\mu\nu}=2{\delta S}/{\delta g^{\mu\nu}},
$
 we find
\begin{eqnarray}
\T_{\mu\nu}=-\rho\,\nabla_\mu\theta\nabla_\nu\theta
+\frac{\rho}{2}\,g_{\mu\nu}\nabla_\sigma\theta\nabla^\sigma\theta
+g_{\mu\nu}V(\rho).
\label{Bemom}
\end{eqnarray}

This energy--momentum tensor is symmetric,
$\T_{\mu\nu}=\T_{\nu\mu}$, by construction and also manifestly.
 Using the equation of motion (\ref{bareq}),
we see at once that $\T_{\mu\nu}$ is traceless,
$\T^\mu_{\ \mu}=0$,
precisely when $V=c\rho^3$ i.e., when our theory has the
conformal symmetry.
Finally, $\T_{\mu\nu}$ is  conserved,
\begin{eqnarray}
\nabla_\mu\T^{\mu\nu}=0,
\end{eqnarray}
as it follows from general covariance
(i. e. from covariance w. r. t. diffeomorphisms \cite{SOUR}),
and also from the eqns. of motion.

Let us assume that the potential is $V(\rho)=c\rho^3$ so that the
system has conformal symmetry.
To any conformal vector field
${X}=\big({X}^\mu\big)$ on $M$,
$L_{{X}}g_{\mu\nu}=\lambda g_{\mu\nu}$,
we can now associate a conserved current \cite{DHP} on
$M$ by contracting the
energy-momentum tensor
\begin{eqnarray}
k^\mu=\T^{\mu}_{\ \nu}X^\nu.
\label{conscur}
\end{eqnarray}
In fact,
$\nabla_{\mu}k^\mu
=
(\nabla_\mu\T^{\mu}_{\ \nu}){X}^\nu
+
\frac{1}{2}L_{{X}}g_{\mu\nu}\T^{\mu\nu}=0.
$
The first term here vanishes
beacause $\T_{\mu\nu}$ is conserved, and the second term
vanishes because $\T_{\mu\nu}$ is traceless.

Let us assume henceforth that the fields $\rho$ and $\theta$
are also equivariant.
Then the Bargmann--space energy--momentum tensor $\T_{\mu\nu}$
becomes $s$-independent.

If  $X^\mu$ commutes with the vertical vector $\xi^\mu$,
one can construct a conserved current on ordinary space out of
$k^\mu$ as follows  \cite{DHP}. $k^\mu$ does not depend on $s$ and
projects therefore into a well-defined current $J^\alpha$ on $Q$,
$(k^\mu)=(J^\alpha, k^s)$.
The projected current is furthermore conserved,
$\nabla_{\alpha}J^\alpha=0$,
because $\xi=\nabla_{s}$ is covariantly constant so that
$\nabla_{s}k^s=0$.
 In the general case, however,
the current $k^\mu$ can not be projected in
ordinary space, because it may depend on $s$;
 $\nabla_{s}k^s$ may also be non-vanishing.

Our idea is to construct a new current out of $k^\mu$ which does
have the required properties.
Let us restrict in fact $k^\mu$ to a ``section''
$s=-\Theta(x,t)$, i.e., define the Bargmann--space vector
\begin{equation}
j^\mu(x,t)=k^\mu\big(x,t,-\Theta(x,t)\big)
=
\Big(\T^\mu_{\ \nu}X^\nu\Big)
\big(x,t,-\Theta(x,t)\big).
\label{projcur}
\end{equation}
Then
\begin{eqnarray}
\nabla_{\mu}j^\mu
=
\nabla_{\alpha}k^\alpha
-\nabla_{\alpha}\Theta\nabla_{s}k^\alpha
=
-\nabla_{s}\Big(\nabla_{x}\Theta\, k^x
+\nabla_{t}\Theta\, k^t+k^s\Big).
\label{kchange}
\end{eqnarray}

Inserting here the explicit form of $k^\mu$ we find that the
bracketed quantity vanishes due to the equations of motion.
The current $j^\mu$ is therefore conserved on $M$,
$
\nabla_{\mu}j^\mu=0.
$
 Let us now define the projected current as
\begin{equation}
J^\alpha(x,t)=
\frac{j^\alpha(x,t)}{\nabla_{s}\theta\big(x,t,-\Theta(x,t)\big)}.
\end{equation}
It can shown using the equations of motion that $J^\alpha$ is a
conserved current on $Q$,
$
\nabla_{\alpha}J^\alpha=0.
$
 Integrating the time--component of the projected current
on ordinary space,
\begin{equation}
\int dx J^t
\equiv
\int\!dx\,\frac{{\cal T}_{\mu\nu}}{\nabla_{s}\theta}
X^\mu\xi^\nu
\equiv
\int\! dx\,\frac{{\cal T}_{\mu s}}{\nabla_{s}\theta}X^\mu,
\end{equation}
is hence conserved for any conformal vector $X=(X^\mu)$.

This yields the same conserved quantities as found before.
The Barg\-mann-space energy--momentum tensor is in fact
related to that in ordinary space, (\ref{BJem}), according to
\begin{equation}
\begin{array}{ccccc}
(\nabla_{s}\theta)\, T_{tt}\hfill&=&-\T^t_{t}\hfill&=
&-\T_{st},\hfill
\\
(\nabla_{s}\theta)\, T_{tx}\hfill&=&\T^t_{x}\hfill&=
&\T_{sx},\hfill
\\
(\nabla_{s}\theta)\, T_{xt}\hfill&=&-\T^x_{t}\hfill&=
&-\T_{xt},\hfill
\\
(\nabla_{s}\theta)\, T_{xx}\hfill&=&\T^x_{x}\hfill&=
&\T_{xx}.\hfill
\label{emtensors}
\end{array}
\end{equation}

Owing to the extra dimension,
the Bargmann--space energy--momentum tensor admits
the new component $\T_{ss}$ which, when contracted with the
``vertical'' component $X^s$ of the lifted vector field, yields
the $-C^0$ term in Noether's theorem (\ref{consquant}).
The situation is nicely illustrated by formulae like
(\ref{Ttransc}) of Section 2.

When $X$ is fiber--preserving, we recover the generators
$H, P, B, N, \Delta, K$ in
(\ref{Galconst})
and  (\ref{newnrconst}) of the Schr\"odinger algebra.

For the non--fiber--preserving vectors,
 we get instead the  new conserved quantities $G, D,
 {\rm C}_1, {\rm C}_{2}$
in (\ref{BJconst}) and (\ref{newrelconst}).

Interchange, $t\leftrightarrow s$, acts on the Lie algebra of
conserved quantities \cite{Duval}. It carries in particular
the energy to particle density, boosts to ``antiboosts'',
etc.,
as already noted in Section 3.

\section{The symmetries of the Schr\"o\-dinger equation}

 We discuss now  the (non--linear) Schr\"odinger
equation in $d$ spatial dimensions,
\begin{equation}
i\p_{t}\Psi
=
-\frac{1}{2}\bigtriangleup\Psi
-\frac{\p\overline{V}(\vert\Psi\vert^2)}{\p\Psi^{*}}.
\label{NLS}
\end{equation}
where $\bigtriangleup$ is the $d$-dimensional Laplacian.
When the wave function is
decomposed into module, $R$, and
phase, $\Theta$,
$
\Psi= R^{1/2}e^{i\Theta},
$
Eqn. (\ref{NLS}) becomes indeed (\ref{eqmotion}), with
\begin{equation}
V=\overline{V}+\frac{1}{8}
\frac{(\p_{i}R)^2}{R}.
\label{effectivepot}
\end{equation}

A non-vanishing effective potential $V$ is obtained therefore
even for the {\it linear} Schr\"odinger equation
$\overline{V}=0$.
The ``free'' theory
described by ${\cal L}_{0}$ in Eqn. (\ref{BJ}) corresponds hence to
a non--linear Schr\"odinger equation (\ref{NLS}) with
effective potential
$\overline{V}=-\frac{1}{8}
\frac{\nabla_{i}R\nabla^{i}R}{R}$, this latter
canceling the term coming from the hydrodynamical
transcription.
As we show below,
canceling this effective term plays a crucial role.

Let us explain everything from the ``Kaluza-Klein type'' viewpoint.
Generalizing to curved space, let us
consider a complex scalar field $\psi$ on a $d+2$ dimensional
``Brinkmann'' space $M$ (\ref{BRINK}).
Generalizing the flat--space results, we posit the action
\begin{equation}
S=
\int\frac{1}{2}\nabla_{\mu}\psi\,\nabla^{\mu}\bar{\psi}
\,\sqrt{-g}\,d^{d+2}x,
\label{BNLS}
\end{equation}
where $g={\rm det}\big(g_{\mu\nu}\big)$. The associated field equation
is the curved--space massless Klein-Gordon
(i.e., the free wave) equation
\begin{equation}
\nabla_{\mu}\nabla^{\mu}\psi=0.
\label{waveq}
\end{equation}

Equation (\ref{waveq}) is {\it not} in general invariant
w. r. t. conformal transformations of $M$,
$f^\star g_{\mu\nu}=\Omega^2g_{\mu\nu}$, implemented
as
\begin{eqnarray}
\psi\to\widetilde{\psi}=\Omega^{d/2}f^*\psi.
\end{eqnarray}
We explain this in the hydrodynamical transcription.
Decomposing
$\psi$ as $\psi=\sqrt{\rho}\,e^{i\theta}$, the action
(\ref{BNLS}) becomes
\begin{equation}
S=
\int\left(\frac{1}{2}\rho\nabla_{\mu}
\theta\nabla^{\mu}\theta
+
\frac{1}{8}\frac{\nabla_{\mu}\rho\nabla^{\mu}\rho}
{\rho}
\right)\sqrt{-g}\,d^{d+2}x.
\label{Baction}
\end{equation}

The action on the fields is now  (\ref{impl}) i.e.
$
\theta\to\widetilde{\theta}=f^*\theta
$,
$
\rho\to\widetilde{\rho}=\Omega^d f^*\rho
$.
As we have seen before,
 the first (``kinetic'') term in (\ref{Baction}) is invariant.
The second term is  {\it not} invariant.
Let us, however, modify the Lagrangian by adding a term
which involves the
scalar curvature ${\cal R}$ of $M$,
\begin{equation}
S_{\cal R}=
\int\underbrace{\Big[
\frac{1}{2}\nabla_{\mu}\psi\nabla^{\mu}\bar{\psi}
+\frac{d}{8(d+1)}{\cal R}\vert\psi\vert^2
\Big]}_{\L_{R}}
\sqrt{-g}\,d^{d+2}x.
\label{BNLSmod}
\end{equation}

Then the symmetry-breaking terms will be absorbed by those
which come
from transforming ${\cal R}$, leaving a mere surface term
(see Appendix C).
In conclusion, the conformal symmetry {\it on M}
is restored by the
inclusion of the scalar curvature term as in Eq.
(\ref{BNLSmod}),
see \cite{CCJ}.
Let us stress that this curvature term is only necessary due
to the
presence of the non-linear potential
$\frac{1}{8}\frac{\nabla_{\mu}\rho\nabla^{\mu}\rho}{\rho}$.
 Restoring the potential, the conformally invariant action is
\begin{equation}
S_{\bar{V}}=
\int\left[\frac{1}{2}\nabla_{\mu}\psi\,\nabla^{\mu}\bar{\psi}
+\frac{d}{8(d+1)}{\cal R}\vert\psi\vert^2
-\overline{V}(\psi^\star\psi)
\right]\sqrt{-g}\,d^{d+2}x.
\label{BVNLS}
\end{equation}

In Minkowski space ${\cal R}\equiv0$. The curvature- term
must nevertheless be added to the Lagrange density, since the
confor\-mal\-ly--transformed metric has already ${\cal R}\neq0$.
The scaling properties of the Lagrangian imply furthermore that
 $\overline{V}(\rho)=c\rho^{1+2/d}$ is the only potential
 consistent
 with the conformal symmetry $\O(d+2,2)$.

So far, we have only considered what happens on extended space.
When the theory is reduced to ordinary space-time, some of the
symmetries will be lost, however. We explain this when $M$ is
$(2+1)$--dimensional Minkowski space and for the linear
Schr\"odinger
equation $\overline{V}=0$.

Firstly, the full conformal group (or its Poincar\'e subgroup)
can only be  projected
to a (field--depen\-dent) action on ordinary space-time
using (\ref{newequiv}) and (\ref{projcond}).
However,
the extended--space model only reduces to
one of the Bazeia-Jackiw form (\ref{BJ}) on $Q$ when the
potential is
$V(\rho)=c/\rho$.
The effective potential in (\ref{Baction})
 is manifestly not of this form, though.
The weak condition (\ref{newequiv}) is
hence inconsistent with the Schr\"odinger equation and has
therefore
to be discarded.

Under the assumption of  equivariance instead,
Eq. (\ref{modequiv}), the wave equation (\ref{waveq})
on Minkowski space reduces,
for $\Psi(x,t)=e^{-is}\psi(x,t,s)$, to the free
Schr\"o\-din\-ger equation
\begin{equation}
i\p_{t}\Psi+\frac{1}{2}\p_{x}^{2}\Psi=0.
\end{equation}
In terms of $R(x,t)$ and $\Theta$ where
$\Psi=\sqrt{R}e^{i\Theta}$,
this equation becomes
\begin{equation}
\begin{array}{c}
\p_{t}R+\p_{x}\big(R\p_{x}\Theta\big)=0,\hfill
\\
\p_t\Theta+\frac{1}{2}(\p_x\Theta)^2=
-\frac{1}{8}\frac{(\p_{x}R)^2}{R^2}+\frac{\p_{x}^2R}{4R}.
\hfill
\label{schroddec}
\end{array}
\end{equation}
(Eqn. (\ref{schroddec}) does not contradict
 (\ref{eqmotion}), since now $V=V(R,\p_{x}R)$).

As explained in Section 5, usual equivariance only allows
the Schr\"odinger subgroup to project~: the ``truly
relativistic''
generators $G$ and $D$, (i.e., the antiboosts and the
time dilations)
as well as conformal generators ${\rm C}_{1}$ and
${\rm C}_{2}$
are hence broken by the reduction, leaving
 us with the mere Schr\"odinger symmetry \cite{JHN},
 \cite{barg}.
This latter is furthermore consistent with the potential
$\overline{V}(\rho)=c\rho^{1+2/d}$.

The conserved quantities can be determined as indicated above.
For the linear Schr\"odinger equation in $(1+1)$ dimensions,
for example,
the conserved energy-momentum tensor (\ref{Schemt})
in Appendix C
allows to calculate the conserved quantities.

{\it On extended space} all conformal transformations are
symmetries,
and (\ref{conscur}) associates a conserved current
$k^\mu(,x,t,s)$,
on $M$, $\nabla_{\mu}k^\mu=0$, to each conformal generator.
Its restriction to
the section $s=-\Theta(x,t)$,
$
j^\mu(x,t)
$
 in (\ref{projcur}), is not in general conserved, though.
In $2+1$-dimensional Minkowski space, for example,
the $\xi$--preserving transformations do yield conserved
currents,
namely
the usual Schr\"odinger conserved quantities \cite{JHN},
\cite{barg}.
However, the currents associated to $\xi$--non--preserving
transformations
as antiboosts, etc. are manifestly not conserved, as seen from
(\ref{kchange}).

Let us conclude our investigations with explaining how the
results of
Jevicki \cite{JEV} fit into our framework.
Let us start with the free wave equation (\ref{waveq}) in $(2+1)$
dimensional  Minkowski space
and let us assume that the scalar field has the form
\begin{equation}
\psi=\frac{1}{4\pi}\big[\Psi(x,t)e^{is}
+\Psi^\dagger(x,t)e^{-is}\big]
=
\frac{1}{2\pi}\sqrt{R(x,t)}\cos(\Theta+s),
\label{jevicki}
\end{equation}
where $\Psi(x,t)=\sqrt{R(x,t)}e^{i\Theta(x,t)}$.
 This field is not equivariant but is rather a mixture of
two states with ``masses'' $(+1)$ and $(-1)$. Hence the usual
theory
of \cite{barg} does not apply. Nor does it fit perfectly
into our ``weaker'' theory~:
the phase is {\it identically} zero, so that any
$\Theta$ solves our equation (\ref{newequiv}).
Calculating the Lagrange density for
the Ansatz (\ref{jevicki}), we find, however,
\begin{equation}
\begin{array}{cc}
-2\pi\L_{0}
=
&\Big\{\2R\big(\p_{x}\Theta\big)^2+R\p_{t}\Theta\Big\}
\sin^2(\Theta+s)
+
\displaystyle{\frac{(\p_{x}R)^2}{8R}
\cos^2(\Theta+s)}\hfill
\ccr
&-
\Big\{R\p_{x}\Theta+\p_{t}R\Big\}\sin(\Theta+s)
\cos(\Theta+s).
\hfill
\end{array}
\end{equation}

The vertical direction can be compactified with period $2\pi$. Then
integrating over $s$  yields the reduced action on ordinary
space--time
\begin{equation}
-\int dxdt
\left[\Big\{\2R\big(\p_{x}\Theta\big)^2
+R\p_{t}\Theta\Big\}
+
\frac{(\p_{x}R)^2}{8R}\right].
\end{equation}
Removing the effective potential $\frac{(\p_{x}R)^2}{8R}$,
 we end up
 with the expression in (\ref{BJ}). It has therefore the same
$\O(3,2)$ conformal symmetry.

\goodbreak
\kikezd{Acknowledgements}.
We are indebted to C.~Duval for sending us his
unpublished notes \cite{Duval} and for many enlightening
discussions.
We would like to thank also
D.~Bazeia, R.~Jackiw, A.~Jevicki, and N.~Mohameddi.
M.~H.  acknowledges
the {\it Laboratoire de Math\'ema\-thi\-ques et de Physique
Th\'eorique}
of Tours University for hospitality, and
the French Government for a doctoral scholarship.

\goodbreak

\vfill\eject

\section{Appendix A~: Lie algebra structure}

The conserved quantities
in (\ref{Galconst}), (\ref{BJconst}),
(\ref{newnrconst}) and (\ref{newrelconst}) form a closed algebra.
The Poisson brackets,
$
\big\{M, N\big\}=
\displaystyle\int\left(\frac{\delta M}{\delta R}
\frac{\delta N}{\delta\Theta}-\frac{\delta M}{\delta\Theta}
\frac{\delta N}{\delta R}\right)dx,
$
read
\begin{equation}
\begin{array}{ccc}
\big\{H,P\big\}=0,\hfill
&\big\{H,N\big\}=0,\hfill
&\big\{H,B\big\}=P,\hfill
\\
\big\{H,\Delta\big\}=H,\hfill
&\big\{H,K\big\}=2\Delta,\hfill
&\big\{H,D\big\}=H,\hfill
\\
\big\{H,G\big\}=0,\hfill
&\big\{H,{\rm C}_{1}\big\}={0},\quad\hfill
&\big\{H,{\rm C}_{2}\big\}=G,\hfill
\\
\big\{P,N\big\}=0,\hfill
&\big\{P,B\big\}=-N,\hfill
&\big\{P,\Delta\big\}=\frac{1}{2}P,\quad\hfill
\\
\big\{P,K\big\}=B,\hfill
&\big\{P,D\big\}={0},\hfill
&\big\{P,G\big\}=H,\hfill
\\
\big\{P,{\rm C}_{1}\big\}=G,\quad\hfill
&\big\{P,{\rm C}_{2}\big\}=2\Delta-D,\hfill
&\big\{N,B\big\}=0,\hfill
\\
\big\{N,\Delta\big\}=0,\hfill
&\big\{N,K\big\}=0,\hfill
&\big\{N,D\big\}=-N,\hfill
\\
\big\{N,G\big\}=-P,\hfill
&\big\{N,{\rm C}_{1}\big\}=2\big(D-\Delta\big),\hfill
&\big\{N,{\rm C}_{2}\big\}=-B,\hfill
\\
\big\{B,\Delta\big\}=-\frac{1}{2}B,\hfill
&\big\{B,K\big\}=0,\hfill
&\big\{B,D\big\}=-B,\hfill
\\
\big\{B,G\big\}=-D,\hfill
&\big\{B,{\rm C}_{1}\big\}={\rm C}_{2},\hfill
&\big\{B,{\rm C}_{2}\big\}=-K,\hfill
\\
\big\{\Delta,K\big\}=K,\hfill
&\big\{\Delta,D\big\}=0,\hfill
&\big\{\Delta,G\big\}=-\frac{1}{2}G,\hfill
\\
\big\{\Delta,{\rm C}_{1}\big\}=0,\hfill
&\big\{\Delta,{\rm C}_{2}\big\}=\frac{1}{2}{\rm C}_{2}
\hfill
&\big\{K,D\big\}=-K,\hfill
\\
\big\{K,G\big\}=-{\rm C}_{2},\hfill
&\big\{K,{\rm C}_{1}\big\}=0,\hfill
&\big\{K,{\rm C}_{2}\big\}=0,\hfill
\\
\big\{D,G\big\}=-G,\hfill
&\big\{D,{\rm C}_{1}\big\}=-{\rm C}_{1},\hfill
&\big\{D,{\rm C}_{2}\big\}=0,\hfill
\\
\big\{G,{\rm C}_{1}\big\}=0,\hfill
&\big\{G,{\rm C}_{2}\big\}={\rm C}_{1},\hfill
&\big\{{\rm C}_{1},{\rm C}_{2}\big\}=0.\hfill
\\
\end{array}
\label{Poissonbrackets}
\end{equation}

In light-cone coordinates, the generators of $\o(3,2)$ acting on
Minkowski space are
\begin{equation}
\begin{array}{cc}
\left.\begin{array}{c}
P_{x}=\partial_{x}
\hfill
\\
P_{0}=\frac{1}{\sqrt{2}}\big(-\p_{t}+\p_{s}\big)
\hfill
\\
P_{y}=\frac{1}{\sqrt{2}}\big(\p_{t}+\p_{s}\big)
\hfill
\end{array}\right\}
\qquad\hfill
&\hbox{translations}
\ccr
\left.\begin{array}{c}
M_{01}=
\frac{1}{\sqrt{2}}\big[(t-s)\p_{x}-x(-\p_{t}+\p_{s})\big]
\hfill
\\
M_{02}=
-s\p_{s}+t\p_{t}
\hfill
\\
M_{12}=
\frac{1}{\sqrt{2}}\big[-(t+s)\p_{x}+x(\p_{t}+\p_{s})\big]
\hfill
\\
\end{array}\right\}
\hfill
&\hbox{Lorentz transf.}
\ccr
\left.d=t\p_{t}+s\p_{s}+x\p_{x}
\right.\hfill
&\hbox{relat. dilatation}
\ccr
\left.
\begin{array}{c}
K_{0}=\sqrt{2}\left[t^2\p_{t}+x(t-s)\p_{x}
-s^2\p_{s}\right]
-\frac{x^2}{\sqrt{2}}(\p_{s}-\p_{t})
\hfill
\\
K_{1}=xt\p_{t}+(\frac{x^2}{2}-ts)\p_{x}
+xs\p_{s}
\hfill
\\
K_{2}=\sqrt{2}\left[x(t+s)\p_{x}+s^2\p_{s}
+t^2\p_{t}\right]
-\frac{x^2}{\sqrt{2}}(\p_{s}+\p_{t})
\hfill
\\
\end{array}\right\}\quad
\hfill
&\hbox{conf. transf.}
\end{array}
\label{conformal}
\end{equation}

The  $X_{i}$ in (\ref{Bgenerators}) provide just another basis
of this algebra~:
\begin{equation}
\begin{array}{cc}
X_{1}=-P_{x}
\hfill
&\hbox{space translation}\cr
\\
X_{0}=\frac{1}{\sqrt{2}}\big(P_{y}-P_{0}\big)
\qquad\hfill
&\hbox{time translation}
\ccr
X_{2}=-\frac{1}{\sqrt{2}}\big(P_{y}+P_{0}\big)
\qquad\hfill
&\hbox{vertical translation}
\ccr
X_{3}=\frac{1}{\sqrt{2}}\left(M_{01}-M_{12}\right)
\qquad\hfill
&\hbox{galilean boost}
\ccr
X_{4}=\frac{1}{2}\left(M_{02}+d\right)
\qquad\hfill
&\hbox{non-relat. dilation}
\ccr
X_{5}=\frac{1}{2\sqrt{2}}\left(K_{0}+K_{2}\right)
\qquad\hfill
&\hbox{expansion}
\ccr
X_{6}=M_{02}
\qquad\hfill
&\hbox{time dilation}
\ccr
X_{7}=\frac{1}{\sqrt{2}}\big(M_{01}+M_{12}\big)
\qquad\hfill
&\hbox{``antiboosts''}
\ccr
X_{8}=\frac{1}{2\sqrt{2}}\big(K_{0}-K_{2}\big)
\qquad\hfill
&{\rm C}_{1}
\ccr
X_{9}=K_{1}
\qquad\hfill
&{\rm C}_{2}
\end{array}
\label{dictionary}
\end{equation}

The Bargmann-space
transformations constructed above are indeed conformal,
$
f^*g_{\mu\nu}=\Omega^2g_{\mu\nu}.
$
The non--trivial values of the conformal factors are
\goodbreak
\begin{equation}
\Omega\quad=\qquad
\left\{\begin{array}{cc}
e^{\lambda/2}\hfill
&\hbox{non-relat. dilatation}\hfill
\ccr
\displaystyle{\frac{1}{1-\kappa t}}\hfill
&\hbox{expansion}\hfill
\ccr
\displaystyle{\frac{1}{1+\epsilon_{1}s}}\hfill
&{\rm C}_{1}\hfill
\ccr
\displaystyle{\frac{1}{(1-\2\epsilon_{2}x)^2
+\2\epsilon_{2}^2ts}}\qquad\hfill
&{\rm C}_{2}\hfill
\ccr
\end{array}\right.
\label{conffactor}
\end{equation}
The factor $\Omega$ associated to the two non-relativistic
conformal
transformations (dilatations and expansions) depends on $t$ only,
while those associated to ${\rm C}_{1}$ and ${\rm C}_{2}$
also depend on the other variables.

\section{Appendix B~: Implementing on fields}

The field-dependent action
of the two relativistic conformal transformations
${\rm C}_{1}$ and ${\rm C}_{2}$
on space-time is
\begin{equation}
\begin{array}{cc}
\begin{array}{c}
x^\star=
x\big(1+\epsilon_{1}\Theta(x^\star,t^\star)\big),
\hfill
\\
t^\star=t+\2\epsilon_{1}x^2
\big(1+\epsilon_{1}\Theta(x^\star,t^\star)\big),
\hfill
\\
{\Theta}^\star (x,t)
=\displaystyle\frac{\Theta({x}^\star,{t}^\star)}
{1+\epsilon_{1}\Theta({x}^\star,{t}^\star)};\hfill
\\
\end{array}\qquad\hfill
&{\rm C}_{1}\hfill
\\
\ccr
\begin{array}{c}
{x}^\star=\displaystyle{
\frac{x+\epsilon_{2}t\Theta({x}^\star,{t}^\star)}
{1-\2\epsilon_{2}x},}
\hfill
\\
{t}^\star=
\displaystyle{\frac{t+\2\epsilon_{2}^2t^{2}
\Theta({x}^\star,{t}^\star)}
{\big(1-\2\epsilon_{2}x\big)^2},}
\hfill
\\
{\Theta}^\star(x,t)
=
\displaystyle{\frac{\big(1-\2\epsilon_{2}x\big)^2
\Theta({x}^\star,t^\star)}
{1+\2\epsilon_{2}^2\,t
\Theta({x}^\star,{t}^\star)}}.\qquad
\hfill
\\
\end{array}\hfill
&{\rm C}_{2}\hfill
\\
\end{array}
\label{rctransf}
\end{equation}

The formula for ${\rm C}_{1}$ is consistent with
(\ref{expC1}),
since
\begin{eqnarray*}
\frac{1}{1-\epsilon_{1}\Theta^\star(x,t)}=
1+\epsilon_{1}\Theta(x^\star,t^\star).
\end{eqnarray*}

These transformations are implemented on the density field
 $R$ according to
\begin{equation}
\begin{array}{cc}
\begin{array}{c}
{R}^\star(x,t)=
\displaystyle{
\big(1+\epsilon_{1}\Theta({x}^\star,{t}^\star)\big)^4
\frac{R({x}^\star,{t}^\star)}{J_{1}^\star}},
\qquad\hfill
\end{array}\hfill
&{\rm C}_{1}\hfill
\\
\cr
\begin{array}{c}
{R}^\star(x,t)=
\displaystyle{\frac{\big(1+\2\epsilon_{2}^2t
\Theta({x}^\star,{t}^\star)\big)^4}
{\big(1-\2\epsilon_{2}x
\big)^8}\,
\frac{R({x}^\star,{t}^\star)}{J_{2}^\star}},\hfill
\end{array}\hfill
&{\rm C}_{2}\hfill
\end{array}
\label{imppr}
\end{equation}
where the Jacobians are
\begin{equation}
\begin{array}{c}
J_{1}^\star
=
\displaystyle{\frac{1+\epsilon_{1}\Theta}
{1-\epsilon_{1}x\p_{x^\star}\Theta
-\2\epsilon_{1}^2x^2\p_{t^\star}\Theta}},\hfill
\\
\cr
J_{2}^\star=
\displaystyle{\frac{1+\2\epsilon_{2}^{2}t
\Theta
}
{\left(
(1-\2\epsilon_{2}x)^{2}
-\epsilon_{2}t (1-\2\epsilon_{2}x)\partial_{x^\star}
\Theta
-\2\epsilon_{2}^2t^2\partial_{t^\star}\Theta\right)
(1-\2\epsilon_{2}x)^{2}
}
}.
\hfill
\\
\end{array}
\label{jacobian2}
\end{equation}
(In these formulae, $\Theta$ means $\Theta(x^\star,t^\star)$).

\section{Appendix C~: Symmetries of the Schr\"odinger equation}

The potential term (\ref{Baction}) is manifestly {\it not}
invariant,
\begin{eqnarray*}
\frac{1}{8}\frac{\nabla_{\mu}\rho\nabla^{\mu}\rho}{\rho}
\;\to\;
\frac{1}{8}\frac{\nabla_{\mu}\widetilde{\rho}\nabla^{\mu}
\widetilde{\rho}}{\widetilde\rho}
+
\left(\frac{d^2}{8\Omega^2}\Big[
\widetilde\rho\nabla_{\mu}\Omega\nabla^{\mu}\Omega\Big]
-
\frac{d}{4\Omega}\Big[
\nabla_{\mu}\Omega\nabla^{\mu}\widetilde\rho\Big]\right).
\end{eqnarray*}

However, the scalar curvature transforms as
\cite{WALD}
\begin{eqnarray*}
{\cal R}\;\to\;
\Omega^{-2}\Big[{\cal R}
-2(d+1)\Omega^{-1}\nabla_{\mu}\nabla^{\mu}\Omega
+(d+1)(2-d)\Omega^{-2}\nabla_{\mu}\Omega\nabla^{\mu}
\Omega\Big].
\end{eqnarray*}

Therefore,
modifying the Lagrangian by adding a term which involves the
scalar curvature ${\cal R}$ as in Eq. (\ref{BNLSmod}) allows us
to absorbe the
symmetry-breaking terms coming from the potential into those
which come
from transforming ${\cal R}$,
leaving us with a  surface term,
\begin{eqnarray*}
\overline{\L}\,\sqrt{-g}
\quad\to\quad
\overline{\L}\,\sqrt{-g}
\;-\;
\nabla_{\mu}\Big(\frac{d}{4\Omega}\nabla^\mu\Omega
\widetilde{\rho}
\sqrt{-g}\Big).
\end{eqnarray*}

The conserved energy-momentum tensor
for the linear Schr\"odinger equation in $(1+1)$ dimensions
is found as
\begin{equation}
\begin{array}{c}
\T_{\mu\nu}=
\rho\nabla_{\mu}\theta\nabla_{\nu}\theta
-
\displaystyle{\frac{1}{2}}g_{\mu\nu}\rho
\nabla_{\sigma}\theta\nabla^\sigma\theta
\ccr
+\displaystyle{\frac{\rho}{4}}g_{\mu\nu}\nabla_{\sigma}
\theta\nabla^\sigma\theta
+
\displaystyle\frac{1}{4}
\displaystyle{\frac{\nabla_{\mu}\rho\nabla_{\nu}\rho}{\rho}}
-\frac{1}{16}g_{\mu\nu}\displaystyle{
\frac{\nabla_{\sigma}\rho\nabla^\sigma\rho}{\rho}}
\ccr
-\frac{1}{8}\nabla_{\mu}\nabla_{\nu}\rho
+\frac{1}{8}\rho\Big({\cal R}_{\mu\nu}
-\displaystyle\frac{{\cal R}}{4}g_{\mu\nu}\Big).
\label{Schemt}
\end{array}
\end{equation}

The first line here is the free expression $\T_{\mu\nu}^0$ in
(\ref{emtensors});
the second line represents the contribution of the effective
potential;
the last line comes from the curvature term.
Remarkably, this latter term contributes even when
{\it initially} ${\cal R}=0$, since the term
$-\frac{1}{8}\nabla_{\mu}\nabla_{\nu}\rho$
is present even it such case.

\end{document}